\documentclass[acmsmall,authorversion]{acmart}
\usepackage[utf8]{inputenc}
\usepackage[final]{pdfpages}
\usepackage{tabularx}
\usepackage{booktabs}
\usepackage{url}
\usepackage{csquotes}
\usepackage{totpages}
\providecommand\tbl[2]{%
 \footnotesize
  \caption{#1}%
  #2}

\usepackage{pdflscape}
\usepackage{afterpage}

\newcommand{\s}[1]{Section~\ref{#1}}
\newcommand{\f}[1]{Figure~\ref{#1}}
\renewcommand{\t}[1]{Table~\ref{#1}}

\usepackage{acronym}
\acrodef{FFQ}{Food-Frequency Questionnaire}
\acrodef{FR}{Food Record}
\acrodef{WFR}{Weighted Food Record}
\acrodef{24HR}{24-Hour Recall}
\acrodef{OCR}{Optical Character Recognition}
\acrodef{RFPM}{Remote Food Photography Method}
\acrodef{SVM}{Support Vector Machine}
\acrodef{BoW}{Bag-of-Words}
\acrodef{BoF}{Bag-of-Features}

\begin{document}

 \markboth{L. Brenna et al.}{A Survey of Automatic Methods for Nutritional Assessment}

\title{A Survey of Automatic Methods for Nutritional Assessment}

\author{Lars Brenna}
\author{Håvard D. Johansen}
\author{Dag Johansen}
\affiliation{%
  \institution{University of Tromsø - The Arctic University of Norway}
  }

\begin{abstract}
Nutritional assessment is key in order to make decisions about the nature and cause of nutrition related health issues that affect an individual. The systematic process of collecting and interpreting relevant nutrition information, however, is still in its technological infancy. Despite technological advances in storage and analysis of nutritional data, methods for collecting data are largely unchanged over the past two decades. It is well documented that these methods have issues that cause under-reporting. Meanwhile, new developments in wearable biometric logging devices have seen increased traction among individuals. This is sometimes referred to as the \textit{Quantified Self} movement. One part of this movement is the development of technological means for objectively collecting nutritional data.  
Nutritional assessment, however, is about to be heavily impacted by emerging computer science technologies, and this survey provides an overview of promising technology approaches supporting nutritional assessment. Both academic and commercial systems are reviewed and categorized. 
\end{abstract}

\keywords{Human-Computer Interfaces, Nutrition, Lifelogging, Diet recording, Computer Vision, Analytics, Smart Health}

\maketitle

\section{Introduction}
Rates of overweight and obesity are continuing to rise worldwide at an epidemic scale. Besides reducing individual's quality of life, these conditions raise risks for serious health problems that include major chronic diseases such as cardiovascular diseases, cancer, and diabetes \cite{flegal2013association}. Individuals, physicians, scientists, and governments are concerned about the rise of obesity and are trying to remedy the situation. If obesity is dealt with preemptively, before it becomes a problem, individuals can be spared from its health impacts and societies spared from its economic burden~\cite{hammond2010economic}. And, as in many other fields, to manage health we need to measure health. That is, to describe relevant health parameters in objective, quantifiable metrics. In hospitals, a patient's health can be objectively quantified in metrics from data sources including blood values, DNA, x-ray images, EMG, EKG, VO$_2$ uptake, body composition scans, and many more. The procedure necessary to do this, however, is too expensive and too invasive to consider applying preemptively to the population as a whole in a scalable, cost-effective manner. 

 Recent technological developments include relatively inexpensive body metric devices that can be owned and used by private individuals. Examples include smart watches, fitness trackers, and body scales with heart rate and body composition measurement. The developments have also brought advanced scanners and tools to professional athletes and fitness centers, where trained personell can offer services previously reserved for public health institutions. The proliferation of such devices enables quantification of many aspects of the health of regular individuals and of the fitness and performance of athletes in objective, comparable metrics. 





Towards measuring health, the remaining component is to objectively quantify \textit{input}, i.e. dietary intake. 
In epidemiology, this is called \textit{nutritional assessment}. 
Epidemiologists investigate patterns and causes of diseases and other outcomes that affect people. 
Nutritional assessment provides key data for epidemiologists studying all topics connected to the food intake of both populations as a whole, and on the individual level. 
The same methods to collect data are used by the scientists who devise and suggest food and nutritional policies for the public, by sports nutritionists who manage the diet of athletes, by physicians helping chronically ill to manage their diet-related diseases, and by individuals doing self-interventions for their own health development.

The classical methods used to perform nutritional assessment are few but clearly defined, 
and have stayed largely unchanged since their inception~\cite{gibson2005principles}. 
Most existing methods for dietary assessment are based on self-reporting. 
A recognized shortcoming of self-reporting methods for dietary intake is that they tend to have bias that weakens the quality of the data~\cite{schoeller1995limitations}. 
Types of bias range from errors in how well subjects remember what they ate, misreporting due to practicalities and required effort to report, to unwillingness to share certain details of their diet. 
The resulting discrepancies may be very large~\cite{witschi1990short}. 
In fact, data collected by the U.S. National Health and Nutrition Examination Survey (NHANES) in the period 1971--2010 are not physiologically plausible when compared to estimated total energy expenditure~\cite{archer2013validity}. 
Many computer scientists have seen this as an opportunity to contribute to the development of technology-based scientific methods for nutritional assessment. 
While the methods used in nutritional science can be described as various forms of weakly validated self-reporting, methods developed in computer science are typically based on objective measurements from validated devices and sensors. 
A wide array of methods and technologies have been developed, but none of these seems to have gained enough traction to replace the standard methods of self-reporting in epidemiology.

Work done in this field is published in a wide array of publications, including nutritional science, healthcare technology, bioengineering, sensor, computer vision and multimedia journals. 
Two previous surveys focussed on studies on just one method, image based dietary intake reporting~\cite{gemming2015image,boushey2016new}. These surveys are published in nutritional science journals, and both compared studies gathered solely from medical and nutritional publications. They also both compared effort and precision in these studies, with traditional methods as a reference. A third previous survey did a systematic review across multiple technologies, made a taxonomy to classify and describe the different technologies, but focussed on readiness to deploy in low income countries and published in a biology journal~\cite{coates2016scaling}. A fourth, earlier survey covered a broad set of technologies from an epidemiological perspective~\cite{illner2012review}, but did not go into detail on each method nor did it develop a taxonomy to describe them. Its focus was on how new technologies could be used in epidemiological studies.

In contrast, this survey covers a broad set of technologies developed to automate nutritional assessment, with an emphasis on contributions and challenges in computer science. We review and categorize various methods for recording dietary intake and nutritional assessment. 
The methods are divided into subjective and objective measures by the level of user-input needed and by the level of detail the resulting data provides. 
We also discuss the strengths and weaknesses of the surveyed methods in terms of their potential to contribute to epidemiological research. 
Our motivation is to introduce the field of nutritional assessment to the computer science community, 
and to show that this is an important area of research with interesting challenges that remain to be solved.


\section{Nutritional Assessment}\label{s:manual}
Nutritional assessment is the systematic process of collecting and interpreting information about the nature and cause of nutrition related health issues that affect an individual~\cite{url-BDA-web}. 

\citet{gibson2005principles} defines nutritional assessment as: 
\begin{displayquote}
	The interpretation of information from dietary, laboratory, anthropometric and clinical studies.
\end{displayquote}

Some information can only be found through laboratory, anthropometric or clinical studies. 
However, to be able to adjust nutritional status through intervention, it is always necessary to 
study dietary intake. Therefore, this survey focusses on methods for dietary assessment.

Researchers typically study clearly defined cohorts over a limited time period.
Ideally, each meal consumed by each cohort member is recorded in such detail that nutritional contents and amounts can be deduced. 
Other meta-parameters like time of day or location might also be recorded.
The challenge is often how to map recorded information to accurate data about digested macro\footnote{The three primary macronutrients are carbohydrates, protein, and fat.} and micronutrients\footnote{Micronutrients are nutrients that are necessary in small amounts, such as minerals and vitamins.}~\cite{prentice2005macronutrients}.

\subsection{Established methods}

The science of nutritional assessment has developed several standards and methods for data collection.
In this section, we will describe three fundamental methods for individual nutritional assessments that are often used in epidemiological studies~\cite{Bingham1994}. 
We consider these methods \textit{manual} as they rely on mostly human effort to record, process and analyze data. 
A more in-depth description of these and other nutritional assessment methods can be found in the book of \citet{gibson2005principles}, and in the book of \citet{driskell2016nutritional}. 

\begin{figure}
\centering
\includegraphics[width=7.5cm,height=7.5cm,keepaspectratio]{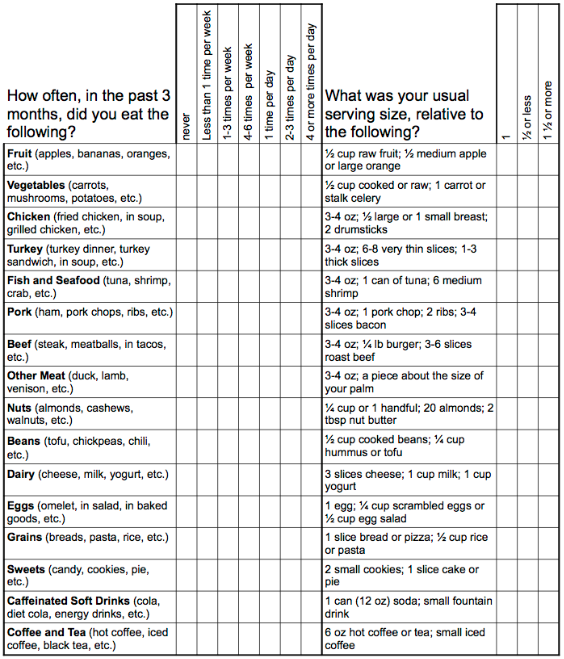}
\caption{A sample pen-and-paper \ac{FFQ} (Source: The European Prospective Investigation of Cancer (EPIC),  http://www.srl.cam.ac.uk/epic/). }	
\label{fig:ffq1}
\end{figure}

\begin{figure}
\centering
\includegraphics[width=7.5cm,height=7.5cm,keepaspectratio]{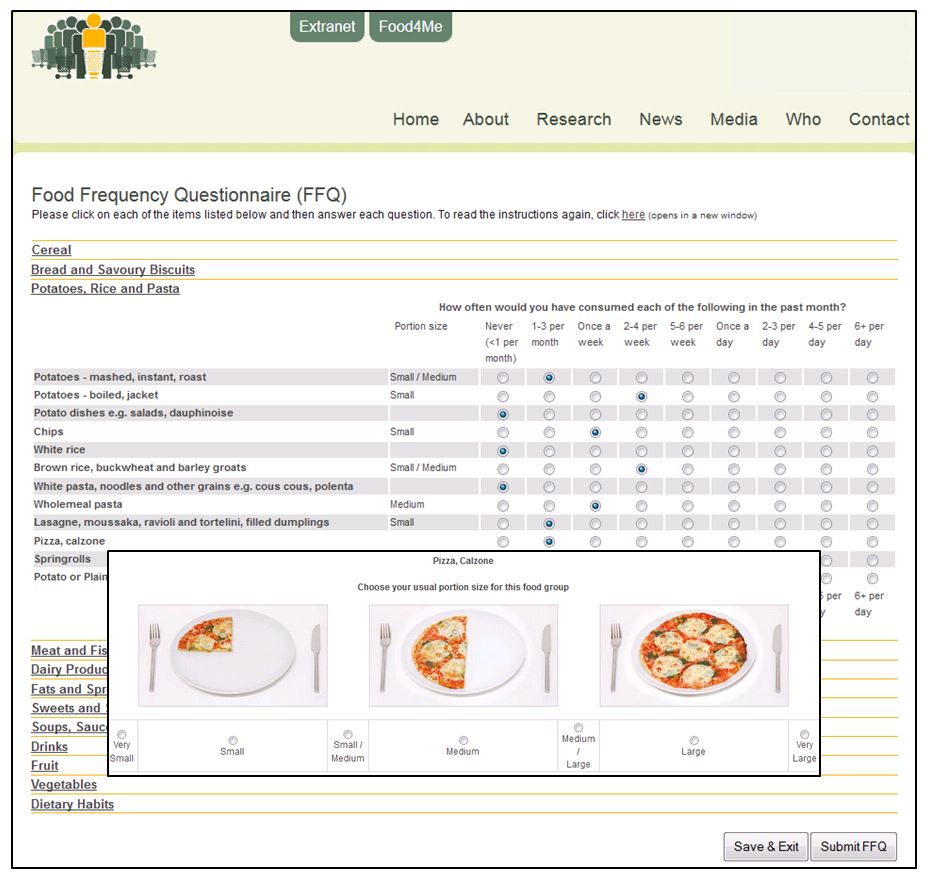}
\caption{A web-based \ac{FFQ} (Source: \cite{forster2014online}). }	
\label{fig:ffq2}
\end{figure}

\paragraph{\ac{FFQ}} 
An \ac{FFQ} is basically a questionnaire designed to record how frequently a subject consumes food and beverage items on a finite list. 
It is typically organized as a table with rows of food and beverage items, and columns indicating frequency and amount for each item.
The subject is then asked to recall from memory how many times they consumed each item in an average day, week or month in the previous year.\footnote{\acp{FFQ} can refer to time periods other than one year, though one year is typical}
Figures~\ref{fig:ffq1} and \ref{fig:ffq2} show two examples of such questionnaires: one pen-and-paper form used by the The European Prospective Investigation of Cancer (EPIC) and one web-based questionnaire. 

The set of items and their associated portion sizes and frequencies vary between studies, depending on the project's research agenda and hypotheses, and on foods typically consumed, which may vary by culture and geography etc. 
Items can be general categories, like fish or meat; or more specific to the needs of the project, like tomatoes or salmon. 
The frequency granularities are limited to only a few options, listed in a less frequent to a more frequent order (e.g., never, less than once a week, and daily).
Portion sizes may be tailored to each specific item (e.g., slices of bread, glass of water).
The period of recall is typically long but limited to, for instance, a year. 
The assumption being that long-term eating habits rarely change significantly, with the exception of dieting for weight-loss etc.

Validity of \acp{FFQ} may be established by cross-referencing results with objective measurements of independent biomarkers in bodily fluids \cite{mckeown2001use}.
Recalling dietary intake from memory can be a challenging task for most people, and filling out extensive questionnaires can take prohibitive amounts of time. 
An \ac{FFQ} with one year recall typically takes a subject one hour to complete. 
It thus is questionable whether subjects are able to give precise data. 
It is also questionable whether \acp{FFQ} are able to account for seasonal or random variations in diet.
These inherent problems are key factors for why \acp{FFQ} are currently designed to only collect low granularity data over long time spans.
The method is still considered to collect an acceptably detailed insight into the nutritional intake of an individual over a longer period of time.

\paragraph{\ac{24HR}}
The \ac{24HR} method is based on structured interviews where subjects are asked about their intake of food and beverages over the previous 24 hours, or from morning to midnight on the previous day. 
The interviewer asks the subject to recall their food and beverage consumption from memory, and typically uses follow-up questions to reveal portion sizes, and how the meals were prepared.
To avoid interview bias, the questions should not be leading.  
In contrast to \ac{FFQ}, where subjects tick boxes of pre-selected items, in \ac{24HR} the interviewer usually has to record the data as free-form text, or as an audio recording.
This flexibility gives finer data granularity and specificity compared to \ac{FFQ}, but puts more strain on the
researchers who need to clean and code unstructured interview data, and then cross-reference identified food items with nutritional tables.
A major issue with \ac{24HR} is diet variability, and the need to capture multiple days to measure consumption of infrequently consumed food items. 
The total amount of work involved, both for the participants and the researchers, makes it difficult to scale \ac{24HR} surveys to large cohorts.

A common practice is to combine \ac{24HR} data with reference data from \ac{FFQ} questionnaires.  
The details acquired from the \ac{24HR} can be used to interpolate the less detailed, but longer spanning data from \acp{FFQ}.


\paragraph{\ac{FR}} 
In the \ac{FR} method, subjects are typically asked to sit down every evening and record what they consumed during that day, in as much detail as possible and perhaps supported by notes taken during the day. 

In contrast to the finite and pre-defined item selection used in \ac{FFQ}, subjects keeping a \ac{FR} are not given a pre-defined item selection. 
In \ac{FR}, data are typically entered as free text records, and sometimes subjects are given forms with fields to fill in. 
If using forms, the schema may determine how portion sizes are described. 
Otherwise, the subject may decide portion size description themselves. 
Thus, the method generally imposes no limits to the variation and specificity of items and amounts which can be recorded. 
It should therefore alleviate the weaknesses of \ac{FFQ} when it comes to random and seasonal variations in diet. 

However, to cover as much details as possible, a recording session will often take more than 15 minutes to complete. 
Due to this high work load, \ac{FR} surveys tends to cover shorter periods of time compared to those based on \ac{FFQ}. 
A typical \ac{FR} survey may only last seven days. 

The challenge of data cleaning and coding also applies to \ac{FR}. Typically, professional dietitians translate and encode the free-text records into schematic data. This can be very time consuming.

\begin{figure}\label{fig:7dfd}
\centering
\includegraphics[width=10cm,keepaspectratio]{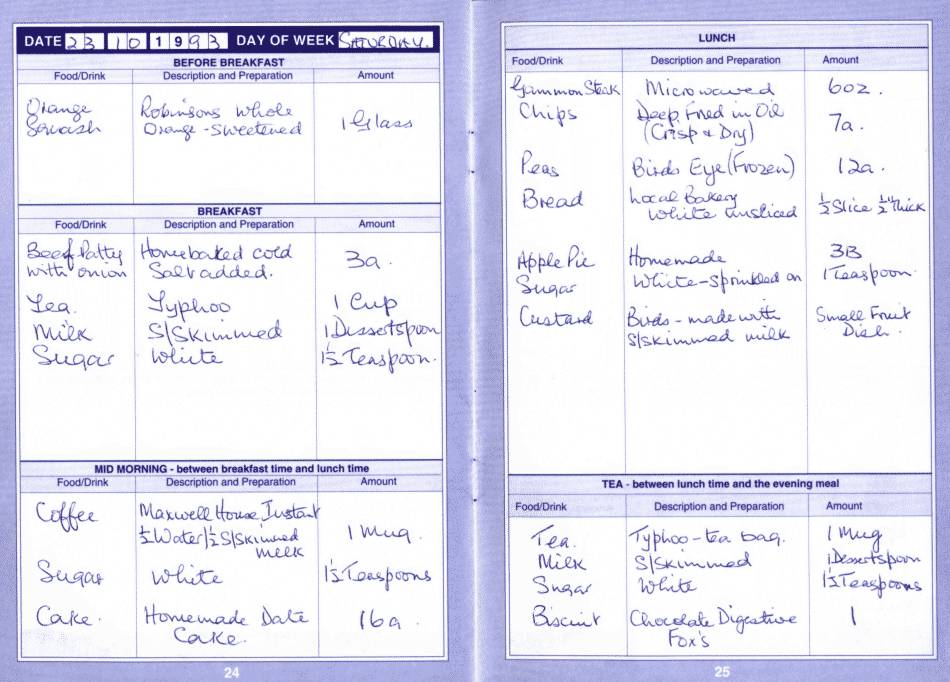}
\caption{A sample pen-and-paper 7 day food record form (Source: The European Prospective Investigation of Cancer (EPIC),  http://www.srl.cam.ac.uk/epic/. ) }	
\end{figure}


A variant aimed to increase the level of recorded details is the \ac{WFR} method, where each subject is equipped with a food scale.
Before each meal, each individual ingredient is weighed on such a scale, and the weight of the ingredient is recorded in the \ac{FR}. 
Alternatively, entire meals can be weighed if it is not possible to weigh the ingredients before composing the meal. 
There are advanced food scales on the marked that come equipped with databases that map food items to its micro nutrient content.

Although food scales can significantly increase the precision of food records on meals composed by the subjects, they are limited to information about ingredients in the food scale's internal database.
The \ac{WFR} method requires higher efforts of the subject than standard \ac{FR}. 
It may not be feasible to bring a food scale around to all places the subject consumes food or liquid. 
In particular, such scales are typically unable to analyze restaurant meals and foods that are already cooked and contain multiple, unknown ingredients.

\subsubsection{Bias}
All three methods are based on subjective self-reporting and have various forms of biases and issues. 
It is likely that the most important bias is that subjects may avoid reporting snacking or under- or over-report amounts. 
Subjects may also not remember details or entire meals. 
A high effort required to report may also reduce adherence to protocol.
The under-reporting bias has been well documented~\cite{witschi1990short}.

\subsection{Classification Parameters}
We will now create a taxonomy that describes the established methods of nutritional assessment. 
The taxonomy and its classification parameters will allow us to compare these methods to new methods suggested by technologists. 

The classification parameters we will use are:
%
%

\begin{description}
\item [Effort.] Dietary intake can be recorded with either explicit or implicit tools. We characterize a method as demanding \textit{minimal effort (min)} if dietary intake recording is an explicit action requiring less than an estimated 30 seconds per session, as \textit{medium effort (med)} if it takes between 30 seconds and three minutes, and as \textit{significant effort (sign)} if it takes more than three minutes. If dietary intake can be recorded implicitly, that is without any explicit action, we characterize the method as demanding \textit{no effort (none)}.
\item [Individualized.] Whether the method can capture the dietary intake of one individual in a group of diners, or whether the method is unable to discriminate between diners.   
\item [Recall.] Whether the method requires that the subject remember past dietary intake, and over how long time.
\item [Schema.] Whether the method lets subjects select food items from lists, express dietary intake in free form, is sensor-based or focussed on recording of meals for subsequent analysis rather than specific foods and drinks.
\item [Portion sizes.] The total dietary intake depends on the amount of each food item. A "Yes" here means that weight or volume is contained in the raw data.
\item [Validity.] Whether nutrient content can be directly derived from raw data and thus be valid for nutritional assessment. To be valid, measured values from the method must be true values according to some gold standard method. If the method is unable to yield true nutrient content values, we characterize it as \textit{invalid}.
\end{description}

In addition, different methods tend to be applied for different durations of time. 
In most cases this is not an inherent characteristic of a given method, so while we have included it in the method descriptions, we have not added duration to our taxonomy.

The traditional methods for nutritional assessments are classified based on our taxonomy in \t{epitable}. 
Since these methods all focus on individual assessment, the \textit{Individualized} category is not included in the table.

\begin{table*}
\centering 
\tbl{\textbf{Manual nutritional assessment methods}}
{
\begin{tabularx}{\textwidth}{llllll}
\toprule
& \textbf{Effort}  
& \textbf{Recall} 
& \textbf{Schema} 
& \textbf{\shortstack[l]{Portion sizes}}
& \textbf{\shortstack[l]{Validity}} 
\\
\midrule
\textbf{\ac{FFQ}} 
& Sign
& \shortstack[l]{Depended on\\memory}
& \shortstack[l]{Food items\\chosen from\\lists}
& \shortstack[l]{Estimated\\from memory}
& Valid
\\[1ex]
\hline
\textbf{\ac{24HR}}
& Sign
& \shortstack[l]{Depended on\\memory}
& \shortstack[l]{Subjects give\\exact food\\eaten}
& \shortstack[l]{Estimated\\from memory}
& Valid 
\\[1ex]
\hline
\textbf{\ac{FR}}
& Sign
& \shortstack[l]{Used at time\\of consumption}
& \shortstack[l]{Subjects give\\exact food\\eaten}
& \shortstack[l]{Weighing\\when\\practically\\feasible}
& Valid 
\\

\bottomrule
\end{tabularx}
}
\label{epitable}
\end{table*}


Significant effort is required for all methods, and particularly \ac{FR} with scales demands a high amount of effort from the subjects. 
The next section will describe how technologists target these issues towards automated nutritional assessment.

\section{Technology for Nutritional Assessment}\label{s:autom}

Much work has been done in trying to modernize nutritional assessment in epidemiology. 
The main goal is to automate all or parts of the recording so that the subject has to spend less time and effort to record data. 
\t{cstable} summarizes the main categories. 
This section contains the surveyed categories and connects them to the taxonomy we developed in \s{s:manual}.

\begin{table}
\centering
\tbl{\textbf{Automated nutritional assessment methods}}
{
\begin{tabularx}{\textwidth}{lllll}
\toprule
\textbf{\shortstack[l]{Device\\type}}
& \textbf{Raw data}
& \textbf{\shortstack[l]{Extracted\\data}}
& \textbf{Strenghts}
& \textbf{Weaknesses} 
\\
\midrule
\textbf{\shortstack[l]{Life-\\loggers}}
& \shortstack[l]{Images\\(GPS, time)}
& \shortstack[l]{Meal, setting,\\ GPS and time}
& \shortstack[l]{Non-invasive,\\passive}
& \shortstack[l]{Low-quality\\images, bad\\angle for food}
\\[1ex]
\hline
\textbf{\shortstack[l]{Digital\\camera\\logging}}
& \shortstack[l]{Images\\(GPS, time)}
& \shortstack[l]{Image of meal,\\metadata}
& \shortstack[l]{High-quality\\images, correlates\\w. time and place}
& \shortstack[l]{Explicit,\\invasive,\\app silos}
\\[1ex]
\hline

\textbf{\shortstack[l]{Mobile\\food\\record}}
& \shortstack[l]{Food record\\recorded on\\smartphone}
& \shortstack[l]{Type of food,\\amount,\\composition,\\image of meal}
& \shortstack[l]{High-quality data,\\high detail,\\correlates w.\\time and place}
& \shortstack[l]{Explicit,\\invasive,\\tedious,\\app silos}
\\[1ex]
\hline
\textbf{\shortstack[l]{Digital\\nutrition\\scale}}
& \shortstack[l]{Weight,\\type of food}
& \shortstack[l]{Calories,\\macronu-\\trients, sugar,\\sodium etc}
& \shortstack[l]{High-quality data,\\detailed}
& \shortstack[l]{Explicit,\\invasive,\\limited to items\\in database}
\\[1ex]
\hline
\textbf{\shortstack[l]{Scanned\\grocery\\receipts}}
& \shortstack[l]{Type of food}
& \shortstack[l]{Calories,\\macronu-\\trients, sugar,\\sodium etc}
& \shortstack[l]{High-quality data,\\detailed}
& \shortstack[l]{Not indi-\\vidualized,\\disconnected\\from meals}
\\[1ex]
\hline

\textbf{\shortstack[l]{Smart\\kitchen\\equipment}}
& \shortstack[l]{Cutting sounds,\\weight changes\\RFID events}
& \shortstack[l]{Type of foods\\being cut}
& \shortstack[l]{Non-invasive,\\passive recording,\\cheap}
& \shortstack[l]{No recording\\of meals}
\\[1ex]
\hline

\textbf{\shortstack[l]{Eating\\gestures}}
& \shortstack[l]{Movements,\\accelerometer}
& \shortstack[l]{Meals,\\activity}
& \shortstack[l]{Non-invasive,\\passive recording}
& \shortstack[l]{No info.\\on nutrition}
\\[1ex]
\hline

\textbf{\shortstack[l]{Chew\\sensors}}
& \shortstack[l]{Chewing\\sounds}
& \shortstack[l]{Meals}
& \shortstack[l]{Non-invasive,\\passive recording,\\cheap}
& \shortstack[l]{Liquids not\\detected,\\no info.\\on nutrition}
\\[1ex]
\hline

\textbf{\shortstack[l]{Swallow\\sensors}}
& \shortstack[l]{Neck muscle\\impedance,\\swallowing}
& \shortstack[l]{Meals}
& \shortstack[l]{Non-invasive,\\passive recording}
& \shortstack[l]{No info.\\on nutrition}
\\[1ex]
\hline

\textbf{\shortstack[l]{Eyeglass-\\integrated\\sensors}}
& \shortstack[l]{Temporalis\\muscle strain,\\accelerometer}
& \shortstack[l]{Meals,\\activity}
& \shortstack[l]{Non-invasive,\\passive recording}
& \shortstack[l]{No info.\\on nutrition}
\\[1ex]
\hline

\textbf{\shortstack[l]{Blood\\flow\\spectro-\\metry}}
& \shortstack[l]{Nutrients,\\accelerometer}
& \shortstack[l]{Nutrients,\\activity}
& \shortstack[l]{Non-invasive,\\passive recording}
& \shortstack[l]{Proprietary\\devices}
\\[1ex]
\hline

\textbf{\shortstack[l]{Bio-\\imped-\\ance}}
& \shortstack[l]{Impedance,\\piezo pressure,\\accelerometer}
& \shortstack[l]{Nutrients,\\hydration,\\activity,\\heart rate,\\sleep and stress}
& \shortstack[l]{Non-invasive,\\passive recording}
& \shortstack[l]{Proprietary\\devices}
\\[1ex]
\hline

\textbf{\shortstack[l]{Lab-on-a-\\chip\\biomarker\\analysis}}
& \shortstack[l]{Biomarkers}
& \shortstack[l]{Nutritional\\status}
& \shortstack[l]{Cheap,\\scalable}
& \shortstack[l]{Limited no.\\of biomarkers,\\no\\macronutrients}
\\[1ex]
\hline

\textbf{\shortstack[l]{Food\\spectro-\\metry}}
& \shortstack[l]{Nutrients}
& \shortstack[l]{Nutrient intake}
& \shortstack[l]{Explicit recording}
& \shortstack[l]{Expensive,\\proprietary\\device}
\\
\bottomrule
\end{tabularx}
}
\label{cstable}
\end{table}


\subsection{Mobile Cameras} 
Collecting every possible data about oneself is often called lifelogging~\cite{gurrin2014lifelogging}. Lifeloggers typically wear one or more body-near devices to collect data as non-invasive as possible about their physiology and the environment during daily life. Effort is usually minimal. The resulting data, in particular images, is valid for use in nutritional assessment, as reported in the works of \citet{gemming2013feasibility,o2013using,thomaz2013feasibility,chen2013using}, and in the survey by \citet{gemming2015image}. The devices used in these studies are worn around the neck, capturing first-person or point-of-view images~\cite{url-sensecam-lifelogger}. Unfortunately, as most meals are presented on a table and the camera is pointing forward, not down, these cameras rarely provide images with good enough detail to practically assess nutritional information. However, they can provide epidemiologists information on meal events and thereby improve recall-based nutritional assessment methods by helping subjects remember all meals, including snacks or drinks. At the time of writing, the only commercially available Lifelogger unit appears to be the Parashoot~\cite{url-parashoot-lifelogger}. 
The \textit{eButton} is an enhanced lifelogging device, which has a stereo camera and includes GPS, a microphone, movement sensors, an audio processor, a proximity sensor, and a barometer~\cite{sun2010wearable,sun2015exploratory}. Estimation of portion sizes is improved using the stereo camera system. The remaining sensors are not harnessed to improve nutritional assessment, so although it is a richer system it shares most of the same characteristics as less advanced lifelogging cameras.


As an improvement over lifelogger cameras, many studies~\cite{reddy2007image,arab2010automated,aizawa2014comparative,Meyers_2015_ICCV} have attempted at doing nutritional assessment by analyzing food images taken with smartphone-based cameras. Conceptually, smartphone cameras and regular digital cameras are equivalent for this method. In contrast to point-of-view cameras, these cameras can be directed and focused directly onto the food, and thereby provide better visual records and a better starting point for analysis. The downside is that while lifelogging devices require no effort, other cameras require explicit effort to record data. 

\subsection{Mobile food record} \ac{FR} as described in \t{epitable} can also be implemented as smartphone-based apps or web apps. Examples include FatSecret~\cite{url-fatsecret-app}, MyFitnessPal \cite{url-myfitnesspal-app}, Fitbit \cite{url-fitbit-app}, Dietist Net~\cite{url-dietitian-app}, Virtuagym~\cite{url-virtuagym-app}, Cronometer~\cite{url-cronometer-app} and SparkPeople~\cite{url-sparkpeople-app}. YouFood is focused on image logging of meals and drinks, rather than detailed text logs with calories and weights \cite{url-youfood-app}. Mobile \ac{FR} is frequently used among athletes and people targeting weight loss or gain. Some apps are community-based, and some include the option of submitting the data to nutritionists. Another frequent feature among the listed examples is to suggest recipes and workout plans based on the goals of the user. 
Mobile \ac{FR} aids recall, usually requires a fixed schema with portion sizes included, and provides valid data. However, the required effort is significant.

\subsection{Digital nutrition scales}\label{ss:scales} Mobile \acp{FR} or \ac{FR} with scales can be augmented with digital nutrition scales to provide more precise records. 
Digital nutrition scales have an internal database of ingredients and their nutritional content.
A digital nutrition scale will therefore not only give exact measurements of portion size, it will also provide immediate nutritional information. 
Examples of scales with built-in screens providing detailed information include:
the NutraTrack~Pro~\cite{url-nutratrack-weight},
the Perfect Portion Weight~\cite{url-perfect-weight},
the Kitrics Digital Scale~\cite{url-kitrics-weight},
the American Weigh NB2-5K-BK~\cite{url-american-weight}, and
the Eatsmart Weight~\cite{url-eatsmart-weight}. 
The SITU Scale~\cite{url-situscale-app} does not have a built-in screen interface, but connects to a smartphone or tablet and offers a rich analysis interface through an accompanying app. 
While the provided data is likely best in class for nutritional assessment, the required effort is significant. 
The scales are limited to information about ingredients in the food scale's internal database and the efforts of the person preparing the meal. 
They are typically unable to analyze restaurant meals and foods that are already cooked and contain multiple, unknown ingredients. 
All scales listed above are readily available commercially. 

\subsection{Scanned grocery receipts} \ac{OCR} can be used on scanned receipts to record food purchases. While restaurant receipts may be ambiguous and difficult to parse for nutritional information, grocery receipts with discrete food items provide clear and valid data with a fixed schema without requiring recall \cite{mankoff2002using}. Using previous shopping history combined with dietary guidelines, \citet{mankoff2002using} show that it is possible to generate product and shopping recommendations. However, the system does not provide individualized data, as it is unable to detect which particular subject is consuming what, whether everything bought is consumed, or the time of consumption. The required effort is unclear, as it likely depends on the method of scanning receipts. 

\subsection{Smart kitchen equipment} Regular kitchen knives equipped with microphones can detect which type of food is being cut~\cite{kojima2016cogknife}. The researchers showed that different types of food have different sound characteristics when being cut, and that this can be used to identify the type.
 Smart fridges registering food items being taken in and out of fridges may accurately identify nutritional information~\cite{gu2009content,luo2009smart}. However, these systems lack the ability to know which food items are eaten, how much and by whom. Thus they do not individualize and do not record portion sizes, so the data is not valid for nutritional assessment of individuals. Effort is minimal and recall is not necessary, as sensors are attached to kitchen equipment that will be used anyway. Currently, no such devices are commercially available. 

\subsection{Eating gestures} Eating usually entails upper body motion. Eating gestures are thus considered a potential source of dietary intake data. Researchers have employed devices ranging from common wrist sensors~\cite{dong2012new,scisco2014examining} to sensor-equipped jackets~\cite{amft2009body} and harnessed surveillance video~\cite{gao2004dining}. The methods are indirect, so to get a measure of how useful bite counts and other eating gestures are for measuring dietary intake, they must be calibrated using a traditional method such as \ac{24HR}. Thus, they can only provide valid data on nutritional content and portion sizes as long as a subjects diet stays unchanged. The advantage is that people are used to wearing watches on their wrist, so the threshold for starting to use wrist-worn bite sensors is likely to be low compared to other methods, and the effort is minimal. Common wrist sensors are commercially available from many vendors, as are cameras for surveillance video. We are not aware of any commercially available jackets with integrated motion sensors such as the prototype used by \citet{amft2009body}.

\subsection{Chew sensors} To non-invasively detect chewing sounds, researchers have placed small microphones on subjects' necks~\cite{sazonov2010automatic,amft2005analysis}. Detecting chewing motion from jaw muscle activation by using a piezo-electric strain gauge sensor has also been investigated~\cite{sazonov2012sensor}. Typically, such sensors can be placed similarly to hearing aids or bluetooth phone headsets, but there may be some threshold for subjects to actually start wearing them. Using chewing sensors to measure dietary intake is an indirect method, as it does not focus directly on the actual food or liquid being chewed, but rather records the sound or motion of chewing. In the earliest studies, the registered data only described dietary intake through identification of meals as events, which does not provide valid data on portion sizes or nutritional content. However, having these data may counteract underreporting in recall-based studies. 

Recently, classification of ingested food from chewing sound analysis has been initiated~\cite{amft2010wearable,passler2012food}. This work could detect certain classes of food texture, like crispy or soft, but not exactly what type nor the amount or weight of food. Classification of drinks and food of drink-like consistency was conjectured to be very hard. If analysis of chewing data can provide valid data on what kind and amount of food is being chewed, then this type of devices may be used for direct nutritional assessment. The reported methods did not require effort from the subject. Currently, no devices for chew detection are commercially available. 

\subsection{Swallow sensors} Another indirect method is to detect swallowing events. This is a field of research that has been on-going since the 1950's. \Citet{Lear:1965aa} and \citet{lear1966swallowing} may have been the first to create swallowing detection apparatus that allowed the subjects to move about relatively freely during sampling. They built two detection systems based on microphones and pneumatic tubes and transducers, respectively. Their systems did have some shortcomings. Pneumatic tubes did not work for subjects that had soft tissue covering the larynx, and the microphones recorded all sorts of background noise that necessitated analysis that was not available at the time. 
Later on, a variety of methods have been applied; electroglottography~\cite{0967-3334-35-5-739}, acoustic sensors~\cite{makeyev2012automatic}, magnetic sensors~\cite{kandori2012simple}, piezoeletric sensors~\cite{kalantarian2015monitoring}, and combining EMG and microphones~\cite{amft2006methods}. The sensors are typically mounted on or near the throat. Although relatively efficient in distinguishing swallowing of food and liquid from other movements and sounds, and perhaps even to detect the amount of food eaten, none of these studies demonstrated any ability to recognize the type of food or nutrition intake. Another problem reported in these studies is that most of these sensors are placed in neck collars, and although they allow free mobility and require no effort to record data, people tended to not like wearing them. Currently, none of these devices are commercially available. 

\subsection{Eyeglass integrated sensors} Rather than making subjects wear completely new devices on new places, some researchers experiment with piggy-backing integrated sensors on existing platforms that many people already use. One such platform is eyeglasses~\cite{farooq2016novel,zhang2016regular}. \cite{farooq2016novel} equipped eyeglasses with EGG sensors to interpret temporalis muscle movements as chewing, and combine that with accelerometers to register physical activity. \Citet{zhang2016regular} limited their study to only use EGG sensors for temporalis muscle movements. None of these studies produced valid data for nutritional assessment and did not suggest how to infer portion sizes or nutritional data. Currently, no such devices are commercially available. 

\subsection{Blood-flow spectrometry} A novel direction is being explored in non-invasive detection of blood values using wrist-worn sensors. Canadian Airo Health claims to have created a wrist sensor that detects nutrients in the bloodstream~\cite{url-getairo-device}. Their device uses a spectrometer to detect nutrients based on different wavelenghts of light. Norwegian Predictor Medical appears to have chosen a comparable approach, with a wrist-worn sensor that at first will detect blood glucose levels~\cite{url-prediktor-device}. Recording data is without effort, the data are individualized, and may provide valid data if their claims are proven. Currently, none of these devices are commercially available. 

\subsection{Bioimpedance} Several consumer market body scales use bioimpedance to estimate body composition. The HEALBE GoBe~\cite{url-healbe-gobe-device} is a fitness tracker and smartwatch that uses bioimpedance to estimate caloric intake from blood glucose variations. While there is limited public information on exactly how it works, the makers claim the wrist-worn device measures fluid flowing in and out of the cells in the body and infers caloric intake from that. While the first generation device received poor reviews of its capability to infer caloric intake, the company started shipping an updated version in 2017 that claims to be significantly improved. The GoBe 2 tracks hydration, calorie intake, calories burned and sleep quality. Recording data is without effort, data are individualized and may be valid for nutritional assessment if the claims are correct.

\subsection{Lab-on-a-chip biomarker analysis}
Laboratory analysis of blood, urine or saliva samples can give measures of nutritional status in vitro without estimation via dietary intake. Typically, lab analysis will give direct or indirect measurements of specific biomarkers that are indicative of nutritional status. While usually very accurate, such analysis is usually expensive, resource-consuming and demands trained personnel. Therefore, it is normally not feasible to use biomarker analysis on large cohorts, or even smaller cohorts that require close tracking outside of hospital settings. 

Lab-on-a-chip solutions seek to alleviate this issue, particularly for point of care locations. \citet{lee2016personalized} and \citet{srinivasan2017precision} provide surveys that cover technological advances in the field, and which biomarkers that can currently be detected. Being able to measure specific biomarkers in a non-expensive and simple manner gives new possibilities for health practitioners at point of care locations outside hospitals, often in rural areas of low-income countries. 

An important facet of point of care measurements is connectivity with electronic patient journals or other online systems for diagnosis and/or tracking of biomarker values. The NutriPhone project~\cite{lee2016nutriphone} is investigating whether biomarker analysis can be made scalable and inexpensive by using smartphone accessory and an app to get readings from one single drop of blood. 
At this point, the list of biomarkers that can be detected is limited. Furthermore, the analysis cannot answer what exactly the subjects have eaten, nor the composition of the food on a macronutrient level. Thus, currently this method does not provide valid data for nutritional assessment.

\subsection{Food spectrometry} Another novel direction is near-infrared spectrometry of food items. Tellspec~\cite{url-tellspec-device} is a handheld device that can provide detailed properties of scanned food items, within seconds. The exact properties returned might vary between food types, but there is currently no public documentation of this. In the case of, for instance, an apple, the device will present its presumed calories, fats, carbohydrates, proteins, fiber, sugar, Titratable acidity (if needed), Brix,\footnote{Degrees Brix (symbol {$^{\circ}$}Bx) is the sugar content of an aqueous solution.} PH, and firmness. Tellspec devices started shipping in 2017.

The SCiO food scanner by Consumer Physics~\cite{url-scio-device} is similar to the Tellspec device. In addition to having a custom device that connects to a smartphone via Bluetooth, Consumer Physics also offers embeddable sensors and a special smartphone with an embedded spectrometry sensor. SCiO devices started shipping in 2017. 

A third food scanner device is from Finnish company Spectral Engines~\cite{url-spectralengines-device}. According to the information on its web page, the device has equal capabilities as the two former devices. It does not appear to be shipping as of the time of writing. 

The Fraunhofer Institute for Factory Operation and Automation IFF claims to have developed an app that uses regular smartphone cameras to detect the constituents of a food item \cite{url-hawkspex-app}. While designated spectrography devices typically use hyperspectral cameras, smartphones are typically equipped with three-channel (red-green-blue) color sensors. The Fraunhofer IFF claims to have overcome the need for expensive and specialized hardware by using the screen to illuminate objects in narrowband colors while simultaneously recording with a built-in (regular) camera. This app is currently a prototype in search of a commercial partner. As of writing, they provide no results, tests, or dates for public availability.

Food spectrometry systems may show promise in providing valid data if they work as promised and the subjects use them correctly and on everything they eat and drink. There is some effort involved, since data recording must happen explicitly. This might work in disfavor of recall, unless used in combination with another method that can record with less effort, for example digital camera logging.

\subsection{Hybrid systems}
Some researchers seek to combine several device technologies in order to gather a more complete record of dietary intake without being too invasive. 

\subsubsection{Wearable ingestion monitors}\Citet{amft2009body} postulate that no single sensor can capture all dimensions of eating behavior. Instead, they combine recording of intake gestures using a motion-sensor equipped jacket, chewing sounds using an ear-attached microphone, with swallowing sensors integrated in a collar. \Citeauthor{amft2009body} also discussed using sensors to detect gastric activity, the thermic effect of food intake, immediate body weight changes, cardiac responses, and immediate changes in body composition. The authors did not present a framework for correlation of composite results from the different sensing systems. 

A similar system suggested by \citet{fontana2014automatic} includes a jaw motion sensor for chewing detection, a wrist-worn proximity sensor that detects motion of the hand and wrist towards the mouth, and an accelerometer to monitor body motion. It was able to non-invasively detect more than 80\% of food intake events, that is, without relying on self-reporting. However, while the work demanded development of novel signal processing and pattern recognition methodologies, the system was not able to classify what food items were actually ingested. 

None of these studies quantified how much the combination of methods improved nutritional assessment vs using just one method. Further work is necessary to produce valid data for nutritional assessment based on multiple sources.

\subsubsection{Lifelogging with food records} Experiments with SenseCam~\cite{url-sensecam-lifelogger} have suggested that lifelogging cameras set to automatically capture pictures can identify meals with high accuracy, but that they do not provide pictures of sufficient quality to derive the content captured  meals~\cite{thomaz2013feasibility,chen2013using}. However, several studies have shown that wearable cameras do improve recall and validity when used together with \acp{FR} or \ac{24HR}. \Citet{o2013using} reported 10 to 17\% higher total caloric intake when pictures was used together with food records versus food records alone. \Citet{gemming2013feasibility} reported 12\% higher total caloric intake when pictures were used together with \ac{24HR} versus without pictures. In both studies, the pictures from the SenseCam provided information on portion sizes, snacks, brand names and forgotten food items. \Citet{arab2011feasibility} reported that 13 of 14 respondents expressed that using a wearable camera was helpful when used to aid memory in a \ac{24HR}.

\subsubsection{Sensor-equipped kitchens} The smart kitchen suggested by \citet{chen2010smart} uses cameras, weights and displays installed in the kitchen counter to calculate nutritional information during the preparation of meals. The subjects expressed great enthusiasm for the support the system offers in making healthy meals. While the smart kitchen may provide nutritional information with high validity and minimal effort, it does not capture who eats what (individualization) or portion sizes. 

In the same category as the smart kitchen, researchers have built a diet-aware dining table with RFID sensors and weights \cite{chang2006diet}. The layered design of the table enables tracking and continuous weighing of RFID-tagged plates being moved around on the table. This lets the researchers track individual dietary intake of each of the items that are placed on the dining table. As such, enhancing the table where food and drinks are placed during the meal improves upon the smart kitchen. However, as with the smart kitchen, it is not feasible to get valid data covering all meals unless the subjects eat all their meals on a smart table.

A more mobile smart-kitchen device currently being developed is the SmartPlate TopView \cite{url-smartplate-app}. A three-part plate supported by a docking station with three corresponding integrated scales allows subjects to divide their meal into three parts and get individual weights and portion sizes. The plate is accompanied by a smartphone app, and the subject registers the meal by taking a top-view image of the plate. The accompanying app analyses the image and uses image recognition to identify and associate the three food items with the weight measurements. The plate can be carried around all day, and perhaps also used in restaurants upon request. As a result, the subject receives a complete report of portion size and estimated nutritional content. As such, this system is potentially able to provide valid, individualized data with minimal effort.    

\subsubsection{Diabetes diary}
Patients with diabetes type 1 must constantly monitor their blood glucose levels and adjust insulin accordingly. As dietary intake is an important factor affecting blood glucose levels, it may be of help to keep a log of what the patients eat. To help with that, researchers have developed systems such as the Diabetes Diary, which allows patients to enter carbohydrates, insulin, and blood glucose levels \cite{aarsand2015performance}. Other relevant data such as time and date, and if used together with a wearable activity tracker, data on physical activity can also be recorded to provide a more holistic representation of the patient. In a nutritional assessment context, the diabetes diary has the same properties as mobile food records. Adding insulin and blood glucose levels, individual glycemic response may be calculated and recorded per food item and used to generate advice for dietary choices.

\begin{table}
\centering 
\tbl{\textbf{Nutritional assessment methods - compared}}
{
\begin{tabularx}{\textwidth}{lllllll}
\toprule
& \textbf{Effort} 
& \textbf{\shortstack[l]{Individ-\\ualized}}
& \textbf{Recall} 
& \textbf{Schema} 
& \textbf{\shortstack[l]{Portion\\sizes}}
& \textbf{Validity} 
\\ 
\midrule
\textbf{\shortstack[l]{Food\\frequency\\questionnaire}}
& Sign 
& Yes 
& Biased 
& \shortstack[l]{Structured} 
& \shortstack[l]{Poor} 
& Valid 
\\[1ex]

\textbf{\shortstack[l]{24 hour\\recall}}
& Sign 
& Yes 
& Biased 
& \shortstack[l]{Non-structured} 
& \shortstack[l]{Poor} 
& Valid 
\\[1ex]

\textbf{\shortstack[l]{Food\\record}}
& Med 
& Yes 
& Good 
& \shortstack[l]{Non-structured} 
& \shortstack[l]{Poor} 
& Valid 
\\[1ex]

\textbf{\shortstack[l]{Weighed\\food\\record}}
& Sign 
& Yes 
& Good 
& \shortstack[l]{Structured} 
& \shortstack[l]{Yes} 
& Valid 
\\[1ex]

\textbf{\shortstack[l]{Lifelogger\\cameras}} 
& None
& Yes
& \shortstack[l]{Software-\\assisted}
& \shortstack[l]{N/A} 
& \shortstack[l]{No} 
& Invalid
\\[1ex]

\textbf{\shortstack[l]{Digital\\camera\\logging}}
& Min 
& Yes 
& Good 
& N/A 
& \shortstack[l]{Depending\\on camera} 
& Invalid 
\\[1ex]

\textbf{\shortstack[l]{Mobile food\\record}}
& Sign
& Yes 
& Good 
& \shortstack[l]{List-based}
& \shortstack[l]{Yes}  
& Valid
\\[1ex]
\textbf{\shortstack[l]{Digital\\nutrition\\scale}}
& Sign
& Yes 
& Good 
& \shortstack[l]{List-based} 
& \shortstack[l]{Yes}  
& Valid
\\[1ex]
\textbf{\shortstack[l]{Scanned\\grocery\\receipts}}
& Med
& No 
& Good 
& Structured 
& No  
& Valid 
\\[1ex]

\textbf{\shortstack[l]{Chew\\sensors}}
& None 
& Yes 
& Good 
& \shortstack[l]{Structured} 
& \shortstack[l]{Poor} 
& Invalid 
\\[1ex]

\textbf{\shortstack[l]{Smart\\kitchen\\equipment}}
& Min 
& No 
& Good 
& \shortstack[l]{Structured} 
& \shortstack[l]{Poor} 
& Invalid 
\\[1ex]
\textbf{\shortstack[l]{Swallow\\sensors}}
& None 
& Yes 
& Good 
& \shortstack[l]{Structured} 
& \shortstack[l]{Poor} 
& Invalid 
\\[1ex]
\textbf{\shortstack[l]{Eyeglass-\\integrated\\sensors}}
& None 
& Yes 
& Good 
& \shortstack[l]{Structured} 
& \shortstack[l]{Poor} 
& Invalid 
\\[1ex]
\textbf{\shortstack[l]{Eating\\gestures}}
& None 
& Yes 
& Good 
& \shortstack[l]{Structured} 
& \shortstack[l]{Poor} 
& Invalid 
\\[1ex]
\textbf{\shortstack[l]{Blood\\flow\\spectrometry}}
& None 
& Yes 
& Good 
& \shortstack[l]{Structured} 
& \shortstack[l]{No} 
& Unknown 
\\

\textbf{Bioimpedance}
& None 
& Yes 
& Good 
& \shortstack[l]{Structured} 
& \shortstack[l]{No} 
& Invalid 
\\

\textbf{\shortstack[l]{Lab-on-a-chip\\biomarker\\analysis}}
& Med 
& Yes 
& Good 
& \shortstack[l]{Unknown} 
& \shortstack[l]{No} 
& Valid 
\\

\textbf{\shortstack[l]{Food spectrometry}}
& Med 
& Yes 
& Good 
& \shortstack[l]{Unknown} 
& \shortstack[l]{No} 
& Unknown 
\\

\bottomrule
\end{tabularx}
}
\label{cstable_tax}
\end{table}

\subsection{Discussion}
 In traditional nutrition assessment of individuals, the goal is to determine which foods and amounts that a subject consumes. 
 When foods and amounts are known, nutrient intake can be derived.
The taxonomy used in \t{epitable} was developed to describe key properties to achieve that goal. The surveyed methods are listed in \t{cstable}, and was elaborated on in this section. As a summary, \t{cstable_tax} categorizes the surveyed methods within our taxonomy. 

 According to \t{cstable_tax}, the two methods that provide the best data are mobile food records and digital nutrition scale. The two methods are also sometimes combined to get the best from both. However, both methods have a low degree of automation and require explicit input as well as significant effort, which tend to lower adherence over time. 
 
 Most remaining automated methods require no or a minimum of effort to record data. Two automated methods (blood flow spectrometry and bioimpedance) claim to capture structured data on nutrition intake. However, blood flow spectrometry is still on the prototype stage and has not been described in literature. Likewise, the accuracy of bioimpedance measuring of nutrient intake has not been published or documented. The accuracy of food spectrometry when it comes to nutritional information of meals has not been independently verified yet. The current food spectrometry methods do not give information on amount of consumed food. 

The variation of sensors and methods, and the variation in the type of data they collect, make it apparent that no single sensor or method is currently able to capture all details and aspects relevant to nutritional assessment. A solution for automated nutritional assessment might therefore have to consist of a combination of input factors, possibly both implicit and explicit, to a get a holistic representation of an individual. This means that work remains in building automated methods that can effortlessly capture precise data on nutritional information of consumed food.


\section{Raw data processing}\label{s:analysis}
Raw data from eight of the 14 automated methods listed in \s{s:manual} are, as shown in \t{cstable_tax}, not valid for nutritional assessment.\footnote{Validity of data from blood flow spectrometry and food spectrometry is currently unknown.} Consequently, researchers have proposed methods to indirectly derive information on dietary intake from raw data that does not contain any direct measurements of relevant parameters. We will divide this overview into sections based on the types of raw data the different methods collect. We here define \textit{analysis} as processing and refining what we previously defined as invalid data, and thereby creating data that is valid for nutritional assessment.

\subsection{Structured data}
Collecting nutritional data using the manual methods described in \s{s:manual} usually requires a step where free text or unstructured data is coded according to a schema, adding portion sizes, frequency, and other information. Such coding steps might introduce errors or misinterpretations, particularly in the case of free-form text descriptions that may not always fit into or translate well to a strict data schema. 
The gold standard for data collection is thus structured data without bias and in a format that can be directly used in nutritional assessment. 

However, not many methods meet this standard. Of the surveyed methods in this review, only mobile food records and digital nutrition scales fully or partially produce structured data. Mobile food records depend on advanced image recognition to estimate portion sizes, and are unable to record the actual weight of a food item. We will discuss image analysis separately in \s{ss:imageanalysis}.

\begin{figure}
\centering
\includegraphics[width=5cm,height=5cm,keepaspectratio]{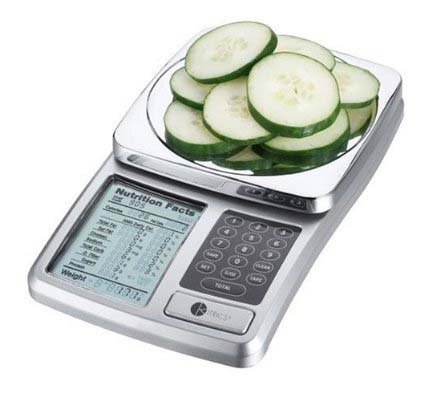}\includegraphics[width=5cm,height=5cm,keepaspectratio]{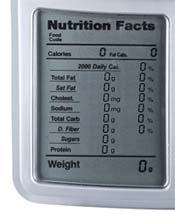}
\caption{Kitrics digital food scale with food packaging label showing macro nutrients and sodium.}	
\label{fig:digscale}
\end{figure}

The digital nutrition scale method described in \s{ss:scales} directly produces structured data with exact portion sizes and nutritional content, and is thus an example of an ideal data producer, at least from an analysts perspective. The schema may vary from vendor to vendor, as may the level of detail when it comes to micronutrients. Some vendors, like the Kitrics model~\cite{url-kitrics-weight} seen in \f{fig:digscale}, has a display that shows the data in the same way as on US food packaging labels. However, it appears that few if any digital nutrition scales have options to export data logs to a central repository or to a remote nutritional consultant.
 
Device data in a structured format lends itself naturally to database storage. In a traditional database-backed application, analysis can be performed through standard SQL interfaces, custom or off-the-shelf analysis software or in a spreadsheet. The SITU scale comes with its own app, and can also be integrated with other apps through a software development kit~\cite{url-situscale-app}. It also allows to share data with, for instance, a physician, coach or nutritionist. 

Food logs recorded in major diet and activity trackers such as Fitbit~\cite{url-fitbit-app} and MyFitnessPal~\cite{url-myfitnesspal-app} can be exported as structured data files. 
These files can then be worked on directly as spreadsheets or imported into a database for analysis. 
The Fitbit app can be used as a mobile food record, and their API enables other systems to fetch records from their database in a structured JSON format.\footnote{https://dev.fitbit.com/docs/food-logging/ (Accessed 27 June 2017)} Their data format is shown in \f{fig:fitbitdata}.

\begin{figure}
\centering
\includegraphics[width=\textwidth,keepaspectratio]{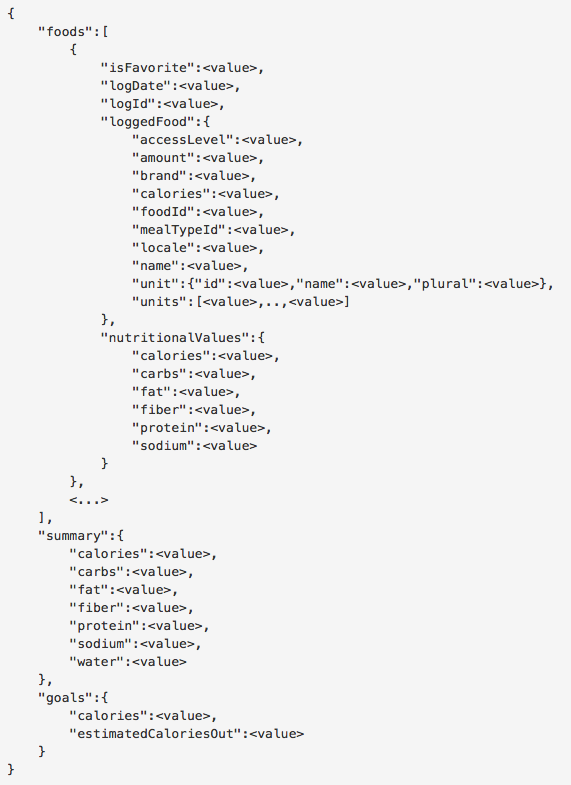}
\caption{Fitbit API JSON data format for food record entries, as of April 2018.}	
\label{fig:fitbitdata}
\end{figure}

Just like medical data in general, having a common data model enables large-scale analysis across data sets from different origins \cite{hripcsak2015observational}. 
However, data formats vary between the studies and applications covered in this survey, and there does not appear to be any standard data format or analysis software/standard specifically for dietary data. 
The Open mHealth initiative aims to make mobile health data more open and accessible~\cite{estrin2010open}. Part of the initiative is to create open schemas for exchange of personal data. Existing schemas include activity, blood glucose, body weight and body temperature, and it may be relevant to implement a standard data format for nutritional assessment as well.

\subsection{Text analysis}
In this survey we cover two categories of textual description of dietary intake. One category is scanned grocery receipts, consisting of structured text. The second category is food records, with its two variants weighed food records and mobile food records. Both types of food records consist of structured or unstructured (free form) text. 

The feasibility of analyzing grocery receipts for nutritional assessment has been the focus of several studies. 
\Citet{martin2006feasibility} recruited shoppers in supermarkets to answer a questionnaire about eating habits and general health aspects, and to volunteer their grocery receipts. 
The resulting data could be described as a form of food records, and were manually transcribed and coded into a form that could be analyzed using non-parametric tests in SAS version 8.\footnote{https://www.sas.com/}
Both the data collection, the data treatment and analysis were laborious. 
Thus, while the resulting data were valid for nutritional assessment, the burden of collection and analysis makes this method impractical for large cohorts.

\Citet{cullen2007food} performed a review of 22 studies that used supermarket sales data to examine food purchasing patterns and thereby assessed the feasibility of using such data for nutritional assessment on a population level. Advantages are that the data is already available in structured form and that the data is objective and free from the biases of self-reporting. Depending on how the data is collected, granularity varies from product amounts sold from a given store to product amounts and frequency sold to specific customers. Most supermarkets already collect and use such data to understand their customers, using sales data registered through loyalty programs and the like~\cite{linoff2011data,chen2012business}.
The collected data from supermarket sales are likely not valid for nutritional assessment of individuals, since it is typically unknown to the system whether (and when) shoppers will actually consume their grocery items, and in households, by whom.

\cite{french2008capturing} performed a review of household food purchasing behavior studies. The review is built on the assumption that to understand the food choices of individuals and how a healthy diet can be promoted, one must study the food purchase behavior of households. This includes collecting grocery receipts, but also receipts from restaurants and take-away meals. It also includes studies that made inventories of food kept in the household at some given point in time. These two methods require manual work to collect, transcribe and code data. 

The third method they investigated was the use of Universal Product Bar Code scanners as a step towards automatization of data collection. However, \citet{french2008capturing} argue that the burden of creating and maintaining a complete product bar code database makes this method impractical and unfeasible. However, multiple studies have since then showed that crowdsourcing can be used to build and maintain a product bar code database, particularly when combined with game play incentives~\cite{budde2010product,dunford2014foodswitch}.
Similar to the work by \citet{martin2006feasibility} and \citet{cullen2007food}, the collected data can be manually coded and analyzed as valid data for nutritional assessment. However, household food purchasing data has the same shortcomings as supermarket sales data in that they do not address individuals.

These studies above are epidemiological studies, using few or no technological aids to automate or systemize the collection of nutritional information. 
In 2008, universal bar code scanning was \textit{"not recommended due to its cost and complexity"}~\cite{french2008capturing}. 
However, in 2018 smartphone cameras are ubiquitous and can be harnessed for grocery receipt scanning. 
There are many apps that can either scan bar codes~\cite{url-groceryiq-app,url-intellilist-app} or use \ac{OCR} to read receipts~\cite{url-evernotescannable-app,url-ibotta-app}. 

Predating the work of \citet{french2008capturing} with six years, \citet{mankoff2002using} presented a system where users who scan their grocery receipts can receive suggestions for healthier shopping alternatives. Using \ac{OCR} they stored all purchased items in a database and compared nutrient intake over time with recommended daily values for each nutrient group. They then applied a simple inference system where deviations greater than ten percent were noted and alternatives suggested. A new shopping list, including suggestions, was presented in printed form.

The manual steps in the system proposed by \citet{mankoff2002using} were clearly imposed by the limitations of technology availability in 2002, and could likely be improved using current smartphone technology. Interestingly, the authors suggest that receipts may have enough information to also estimate what people are actually eating, as opposed to other studies that can only report on what people are purchasing. In this work too, the results are limited due to technological constraints. Their inference algorithms used to predict what people are eating based on their shopping would likely today be replaced by machine learning based algorithms. If successful, this extension would make the data in \cite{mankoff2002using} valid for nutritional assessment of individuals.

\subsection{Image analysis}\label{ss:imageanalysis}
Food descriptions, often accompanied with pictures, are frequently posted in social media channels. With the invent of digital cameras and smartphones, a common behavior is to take pictures of meals and post them online. The willingness and eagerness of people to do this make food images an interesting source of data to determine what people eat.

\begin{figure}
\centering
\includegraphics[height=10cm,keepaspectratio]{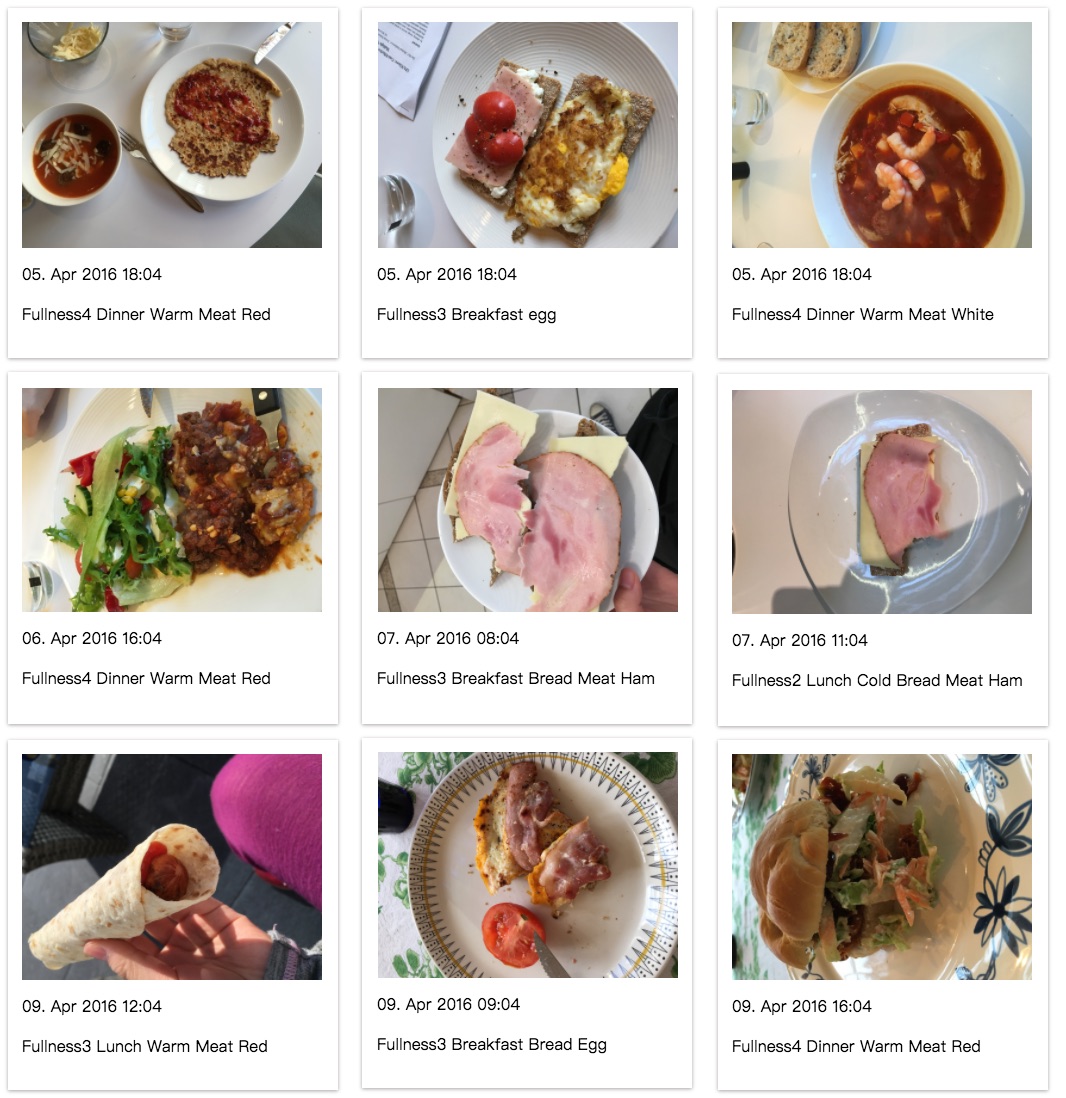}
\caption{Screenshot from a mobile food record prototype using images to record meals. The pictures are accompanied with metadata for time of day and an experimental indicator of fullness (from 1-5). (Credits: Magnus Stenhaug, UiT) }	
\label{fig:magnusdata}
\end{figure}

However, there are significant challenges in the automatic detection of food from images. For instance, there is almost no limit to how food on a plate can be configured. Images can be of individual food items, of full meals with drinks, of plates with clearly separated food items or of plates consisting of multiple components that are mixed together in various degree. Food may also contain ingredients such as sugar, salt, spices, or fats that are not easily captured in an image. This complexity makes detecting food from images a very complex task. 

To simplify the task of food analysis from images it has been suggested to use supplemental metadata describing the meal, such as text or speech. It has also been suggested that putting humans in the loop can be an efficient solution, as a human analyst may apply common knowledge to the analysis and contribute where computer analysis is weak. We call these forms of food analysis from images and metadata \textit{semi-automated image analysis}. However, requiring metadata along with food images adds to the effort required for data capture. With the gradual improvement of object recognition technology using machine learning, this has increasingly been applied to nutritional assessment from images. We call this \textit{automated image analysis}. This section will discuss work in these two areas. 

\subsubsection{Semi-automated image analysis}
Several systems have been suggested where images of subject's meals are seamlessly shared with professional nutritionists who, in turn, estimate the meal's nutritional value. 
This method is commonly referred to as \textit{\acf{RFPM}}~\cite{martin2009novel}. 
The key underlying assumption in these systems is that trained professionals are better at estimating nutritional content of a meal
 than subjects or even algorithmic computer-based assessments.

The seminal work on the \ac{RFPM} approach, was the Wellnavi system~\cite{da2002validity,wang2006development}.
In Wellnavi, subjects use a specifically developed PDA. It has a stylus device, which must be placed next to the food as a reference when an image is taken. Along with the image, the Wellnavi requires that food items and ingredients are listed as text using the stylus. The collected data is then sent to a human dietitian for analysis. 
The dietitian is also able to request that the subjects fill out a short questionnaire if further details are necessary for analysis.  

\begin{figure}\label{fig:wellnavi}
\centering
\includegraphics[width=7cm,height=7cm,keepaspectratio]{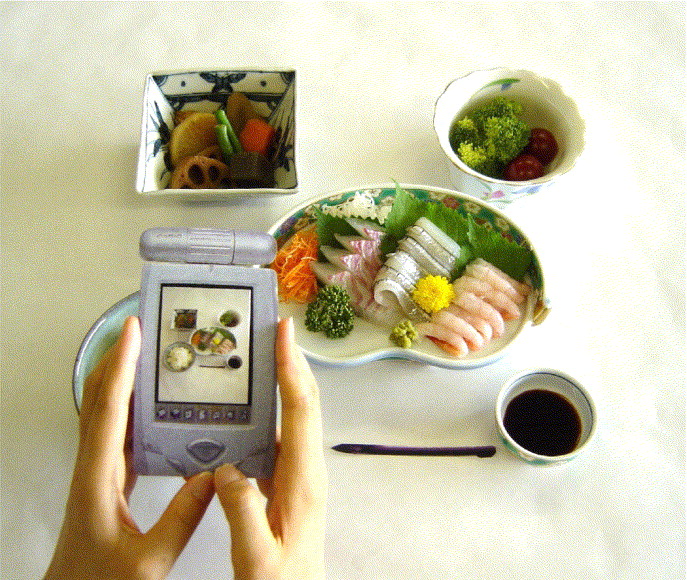}
\caption{The Wellnavi hand-held personal assistant for nutritional assessment (Matsushita Electric Works, Ltd, Osaka, Japan, 2007).}	
\end{figure}

The Wellnavi system was later evaluated against \ac{24HR} and \ac{FR} (with scale) using a more diverse cohort~\cite{kikunaga2007application}. 
That study reported significant differences in nutrient estimation between the three methods. 
The under-reporting bias that Wellnavi intended to alleviate was still found to be present. 
The biggest complaint from the subjects was that they worried what the dietitians, being strangers, would think of their food from the personal images they were sent. 
\Citet{martin2009novel} replicated the Wellnavi studies, with the exception that their system did not automatically send images online to the dietitians. 
However, \citeauthor{martin2009novel} reported good results from nutrient estimations compared to standard methods. 
This suggests choice of cohorts, cultural aspects, and study settings may interfere with reporting.

Placing humans in the computational loop is expensive both financially and computationally, limiting the use of \ac{RFPM} to smaller cohorts.
PlateMate~\cite{noronha11:platemate} tries to alleviate these scalability concerns by decomposing the task of estimating nutritional content into several smaller steps,
and executing them as a system of Mechanical Turk\footnote{Amazon Mechanical Turk can be accessed at www.mturk.com} tasks. 

Neither of these systems give any estimate of the accuracy of professional nutritionists (or whoever executes Mechanical Turk tasks), 
nor do they give any guarantee on the expected delay from when a subject submits an image to when they receive an estimate of the nutrients in the meal they are about to have. 
This may work for some applications, but for someone considering whether or not to eat a meal or parts of a meal on a plate in front of them, 
a significant delay makes the application less valuable.

The Food Intake Visual and Voice Recognizer (FIVR) \cite{puri2009recognition} is functionally similar to the GoCARB system referenced above, but also asks the user to list food items using speech.
This alters the process somewhat, as the segmentation and recognition stage then is designed to identify the listed food items. FIVR also computes 3D volumes of the listed food items.

\subsubsection{Automated image analysis}\label{sss:imageanalysis}
Automated image analysis is often carried out as a multi-step machine learning algorithm, where one first has to segment the parts of the image that contains food and then determine the type and amount of food in each segment. Much work has been done in this area. The surveyors have attempted to cite a representative selection of articles.

In the FoodCam system~\cite{kawano2015foodcam}, the user is asked to give a bounding box around each region of food in a picture, upon which the system will provide a food type suggestion along with associated nutritional information. 
While their system is highly accurate, it depends on a small (learned) set of categories of food frequently found in the geographical vicinity of the researchers. 
Different foods are eaten in different countries, so the system must be re-trained and adapted to work in other localities. 
Their system applied a linear \ac{SVM} to classify food types, achieving a classification precision of 79.2\% in a 100 category food-image dataset.
In a follow-up paper, classification performance was improved using deep convolutional neural network techniques~\cite{yanai2015food}. 
Later, another team used the same datasets and further improved deep convolutional neural networks to achieve a 94.6\% accuracy~\cite{liu2016deepfood}.

\Citet{rich2016towards} analyzed Instagram food images and addressed both images and associated hashtags. They showed that, despite the noisy nature of Instagram hashtags, it is possible to detect various categories of food by relying purely on user labels. This may allow a wider selection of food categories than with pre-defined categories. Their system was implemented using pre-trained convolutional neural networks from the MatConvNet toolbox.\footnote{http://www.vlfeat.org/matconvnet}

\Citet{chen2015saliency} addressed the challenge of automatic food segmentation from images, and proposed to create a salient map of food regions in a given input image. A geometric model is then used to extract the food regions and present them for analysis. Their system estimates portion sizes, but does not attempt to determine food types or to provide nutrient information. 

The Snap-n-Eat system~\cite{zhang2015snap} combines the automated food
segmentation, similar to that by \citeauthor{chen2015saliency}, with estimation of nutritional content, similar to the work of
\Citet{kawano14c}. The images are processed in five steps: saliency detection, segmentation, feature extraction, single food detection, and context reasoning. 
The evaluation of Snap-n-Eat is based on about 2,000 images, manually annotated and categorized in 15 food types, to train a \ac{SVM} for classification. 
Portion sizes per food segment are estimated by counting the pixels in each corresponding segment. 
This simplification assumes that each food category has a fixed nutrition density, and does not take invisible ingredients into account. 
With these assumptions, Snap-n-Eat is able to give a reasonable quick estimate of the nutritional content of a meal with minimal effort as defined in \s{s:manual}. 

The food segmentation step is necessary because a plate can have a large number of configurations and combinations of separate food items or regions. However, different combinations, arrangements and spatial differences of the same food items or ingredients do not necessarily mean two dishes are not the same food. 
This complicates the use of direct image recognition techniques. 
As an alternative, a \ac{BoF} model, which derives from the \ac{BoW} model used in text analysis, has been developed~\cite{fei2005bayesian}. 
In the \ac{BoF} model, an image is represented by a histogram of codewords (or textons), which are defined as representative image patches of commonly occurring visual patterns. 
The image recognition begins by identifying all the codewords in a given image, which is the \ac{BoF} model creation. 
This step is comparative to the segmentation step. 
The second step is classification, as in the other image analysis methods; to find the image category that best matches this distribution of codewords. 
In the classification step, several machine learning methods can be applied.

\Citet{anthimopoulos2014food} applied the BoF model to food images. The training set consisted of nearly 5,000 food images downloaded from the Web and manually categorized. While their system has reasonable accuracy, the BoF model has some errors that other segmentation methods do not. For example, meat is hard to recognize because it may occur on its own, but also as part of other dishes, like soup, pizza or omelets.

Finding the volume of food in a given meal is a related problem to that of identifying the food type. As \citet{dehais2013food} show, it is possible to create a 3D model of a meal on a plate and use that model to estimate its volume. This is increasingly useful as smartphones become equipped with stereo cameras. While their method has promising accuracy, it is questionable whether its computing performance is good enough to provide a realtime-like user experience. Thus work remains in improving the performance of such applications.

The 3D-model method was applied in later work from the same authors, where an application called \textit{GoCARB}~\cite{anthimopoulos2015computer} includes food segmentation, food recognition and volume estimation. GoCARB requires that a credit card-sized reference card is placed next to the meal, before two images are captured with the subjects smartphone in a guided process to ensure best possible image quality. To simplify segmentation and recognition, the meal must be on one elliptical dish where all food items are fully visible. Non-overlapping is important for volume estimation. Food segmentation uses color processing and then region growing and merging, and achieved an accuracy of 88.5\%. Food recognition applies a hierarchical k-means algorithm to histogram-based feature sets, and feeds that to a \ac{SVM}  with a reported 87--90\% accuracy. When food items and their volumes have been computed, the nutrient content of the meal can be drawn from the USDA food composition database.   

The GoCARB system was developed with type 1 diabetes patients in mind, to help them estimate carbohydrate intake in the context of an artificial pancreas.\footnote{Type 1 diabetes patients basically have pancreases that have stopped regulating blood glucose levels. The patients need to measure their blood glucose levels and inject insulin accordingly, which is what the pancreas does in healthy persons. An artificial pancreas consists of a continuous glucose monitor that is attached to the skin and measures blood glucose with short intervals, a subcutaneous insulin injection pump and some sort of control loop algorithm that takes the blood glucose measurements as input and outputs control signals for when and how much insulin should be injected by the pump. Design of the control algorithm is an active research topic on its own, but outside the scope of this survey.}  The glucose monitors used in artificial pancreases usually measure the glucose level subcutaneously. The delay in subcutaneous levels versus levels in the blood can be significant. External factors such as meals, their composition, and physical exercise also affect the bodys insulin need. A delayed or wrongly dosed insulin injection may cause distress or discomfort. It is therefore helpful to be able to better predict blood glucose levels for better timing and dosing of exogenous insulin, and \citet{agianniotis2015gocarb} showed that using the GoCARB system together with an artificial pancreas increased control of insulin dosage and blood glucose levels. 

The task of recognizing food items can be simplified by reducing the problem to recognizing a dish from a list of known menu items where pictures of the dishes have been processed beforehand \cite{beijbom2015menu,bettadapura2015leveraging}. Nutrient information and portion size are then typically known and can be selected from a database. \Citet{Meyers_2015_ICCV} showed that by applying a convolutional neural network this method scaled to 23 restaurants and 2,571 menu items. This might work for individuals whose food habits consist in variations on a limited number of dishes from local restaurants. However, it is not feasible to work on a global scale or with inconsistent food choices, such as home-made food that does not look the same every time. However, \citeauthor{Meyers_2015_ICCV} showed that the same method could be extended to identify food segments, so that their system could be used in the wild. Another interesting contribution from the works of \citet{Meyers_2015_ICCV} was volume estimation from a single image using a convolutional neural network. 

\subsection{Image recipe recognition}
Analysis of text as well as images for nutritional content both have weaknesses and strengths. Given that recipes should contain ingredients that are invisible in images, combining recipes with images should give more nutritional information than images alone. However, only recently have a few data sets been published to allow research on this combination. \Citet{wang2015recipe} were the first to include both recipes and images in one dataset of about 100,000 recipes with one image each. The authors demonstrated a search engine where queries containing images would return a list of the most relevant recipes. Convolutional neural nets were used for image recognition, and word2vec~\cite{mikolov2013efficient} for ingredient extraction from recipes. However, the recipes are represented as raw HTML which makes ingredient extraction more difficult.
A larger dataset was presented a year later by \citet{chen2016deep}, but included recipes and images of Chinese cuisine only. 

The largest and most generalized dataset currently available is the Recipe1M\footnote{http://im2recipe.csail.mit.edu}, containing one million recipes and 800,000 images \cite{cSalvadorc}. Machine learning algorithms to connect recipes and images are complex. The work of \citet{cSalvadorc} share some methods with the work of \citet{wang2015recipe} and the work of \citet{chen2016deep}, but introduces neural joint embedding models that achieve impressive performance. The use of deep convolutional neural networks for image
representation and the use of word2vec vectors~\cite{mikolov2013efficient} for ingredient lists are common for all these works. Their methods are similar to those described in Section~\ref{sss:imageanalysis}. 

 Recipes have two major components: ingredients and instructions. \Citet{cSalvadorc} uses a bi-directional long short-term memory to extract actual ingredient names from ingredient text. For instance, in "3 cups of cream" the \textit{cream} is extracted as the ingredient name and used as a single word in word2vec. 
 \Citet{cSalvadorc} uses one test scenario that is similar to that of \citet{wang2015recipe}, where a query retrieves a recipe from collection of test recipes. In 65\%, their system retrieved a correct recipe. Secondly; they compared their performance with human performance using Mechanical Turk tasks. When challenged choosing the one correct recipe out of 10 provided recipes, for a given food image, humans performed 6.8\% lower than their joint embedding model.

The weaknesses of image recipe recognition are similar to those of automated image analysis, in that invisible ingredients and homogenized food items (such as smoothies, soups, and stews) are challenging to recognize.

 
\subsection{Sound analysis}
Sensor systems designed to capture sounds of eating and detect swallowing and/or chewing generally need to perform two tasks: 1) separating food intake sounds from other sounds, and 2) identifying the type of food being consumed. All surveyed systems use some form of machine learning for these tasks.

CogKnife used sound from a microphone attached to a knife, but in this case it is safer to assume that most cutting sounds will be from cutting food so the challenge can be limited to identify the type of food being cut~\cite{kojima2016cogknife}. The authors extract spectrograms of the cutting sounds and use them as feature vectors to train a machine learning classifier, testing both a \ac{SVM} and a convolutional neural network for comparison.   

\Citet{amft2005analysis} identified that chewing sound analysis could be used as one component of a suggested set of techniques to automate nutritional assessment. 
The set of techniques they suggested included: 1) wearable sensors to detect chewing, swallowing and hand motions, 2) environmental augmentation (i.e., smart kitchen equipment), and 3) high-level context detection. Focussing on chewing sound analysis, they applied sound analysis to separate chewing from speech, isolated and partitioned individual chews and used a decision tree classifier to classify food products against speech signals~\cite{witten2005data}.
Acoustical methods have relatively low accuracy. For example, \citet{sazonov2010automatic} reported an accuracy of 84.7\% using trained \acp{SVM} while 
\citet{amft2005analysis} reported between 80 and 100\% on four different types of food.

\subsection{Motion analysis}
The act of eating and drinking normally involves moving certain limbs, such as hands, arms, jaw and larynx. The challenges are similar to those in sound analysis. The challenge in motion analysis is generally to detect and differentiate eating gestures and motion from speaking and other types of motion.

To detect large upper-body motions like hand-to-mouth researchers have employed devices ranging from common wrist sensors \cite{dong2012new,scisco2014examining} to sensor-equipped jackets \cite{amft2009body}. The metric used is "bite count", as in the number of times the subject has taken a bite of food. \Citet{dong2012new} defined a bite as a certain wrist rotation motion, and was able to detect this using a gyroscope. With this rather simple definition, they could use data from the device without any form of machine learning algorithms for pattern recognition. Precision when compared to manually annotated video recordings was above 80\%. One limitation of a wrist-worn bite counter is that it needs to be started and stopped for each meal. \Citet{amft2009body} did not want this limitation, and used machine-learning techniques to detect and distinguish eating gestures from other natural movements. This work contain few actual details, but mentions using a Naïve Bayes classifier preceded by linear discriminate feature extraction.

\Citet{sazonov2012sensor} used a jaw-mounted piezo-electric strain gauge sensor to detect chewing motion, and achieved 80.98\% accuracy in 
detecting and classifying chewing from other types of lower jaw motion.
The authors used \acp{SVM} trained with the most relevant features as selected in a forward feature selection.
In later work, \citet{farooq2016novel} deteced temporalis muscle strain using a piezoelectric strain sensor. They compared two approaches to differentiate eating from not eating. The first approach used a single classifier with sensor fusion, the second used two separate linear \acp{SVM} trained for eating and activity detection. Combined using a decision tree, an accuracy of 99\% where achieved.

\Citet{zhang2016regular} monitored chewing using eyeglasses with integrated EMG and skull vibration sensors. 
EMG data are passed through a bandpass filter, processed to obtain envelope and divided into sliding windows of 200\,ms. 
Statistical analysis was then used to detect chewing. 

By analyzed data from an EGG device attached to the throat,
\citet{0967-3334-35-5-739} show that swallowing can be detected with
an accuracy of 90.1\%, compared to an accuracy of 83.1\% using an
acoustic method. A three-layer feed-forward neural network was
used to distinguish and detect swallowing from speech, head movements
and other bodily sounds.

\subsection{Spectrography analysis}
The spectography analysis methods of blood and food that are cited in this survey appear to be proprietary and are to our knowledge not published.


\section{Summary of fundamental properties}
In \t{cstable} and \t{cstable_tax}, we established a taxonomy across six fundamental properties of automated systems for nutritional assessment. 
We will next use this taxonomy to discuss how the surveyed systems in general meet these properties, and to suggest future work to improve.  

\subsection{Effort}
It is generally an assumption that to reduce bias, self-reporting should be replaced or supported by systems that alleviate the effort of reporting dietary intake. The surveyed methods that have been categorized as requiring no effort all use some sort of wearable sensors. Many of the studies have reported that wearing the sensors were considered uncomfortable or annoying, some to such a degree that they would not wear them voluntarily outside the study. 

This may suggest that there is a need to develop sensors and devices that are even smaller and can be worn without annoying the subject.

\subsection{Individualized}
Only two of the surveyed methods did not produce data that could immediately be associated with a specific individual. For the two remaining methods, smart kitchen equipment and scanned grocery receipts, further work in signal processing and pattern recognition might be able to reveal distinct patterns that could be used to identify individuals. Smart kitchen equipment could potentially learn how to recognize certain motion or eating patterns to differentiate between subjects. It was also suggested by the authors that over time, it could be possible to extract patterns and differentiations between individuals based on grocery receipts for a family or group. 
This may suggest that further work is needed in signal processing and pattern recognition for individualization. 


\subsection{Recall}
The recall in all surveyed methods is good, meaning there are no documented methods where there is a significant chance of missing events. However, mobile food records and other methods using digital camera logging have been shown to have the same self-reporting bias as traditional methods. This was particularly apparent in those methods that used human dietitians to analyse food images. 

These findings suggest that for image-based nutritional assessment to work, it is essential to continue improving machine-based detection of food items and portion size. 

\subsection{Schema}
Most of the surveyed studies are domain-specific prototypes developed for a limited audience. As such, none of these have emphasized the schema used to store and format recorded data. The surveyors have not been able to find any suitable open source schemas that could have been applicable. A common schema would simplify comparison between methods, and ensure that data can be compared and mined across services. 

We therefore recommend an initiative to form a standard schema for automated nutritional assessment. Open mHealth initiative~\cite{estrin2010open} may be a suitable platform for that.

\subsection{Portion sizes}
Very few of the surveyed methods are able to automatically estimate portion sizes. Sensors for chewing and swallowing are conjectured to do so, pending further work in signal processing and pattern recognition. The method closest to a reliably working system appears to be 3D-based volume estimations from cameras. The surveyed methods include using reference objects in the image and taking several images with different angles. However, recent smartphones have started to include 3D-cameras, most likely to enable augmented reality applications. 

We therefore recommend studies that include 3D-cameras and augmented reality to estimate portion sizes. The gold standard would be real-time feedback to the subject regarding portion sizes according to some diet goal. 

\subsection{Validity}
Four out of 14 surveyed methods were categorized as providing data that is valid for nutritional assessment. This means that very few methods actually produce data that can directly be used to estimate types and amounts of nutrients in meals. Getting valid data appears to be a hard problem, particularly when using a method that requires low effort. 

A low-effort method producing valid data must likely be a combination of methods. For example, one can imagine a smartphone application that combines image recognition for food item identification, 3D-scanning for volume estimation, and spectrography for nutrient information.   


\section{Considerations from an engineering perspective}
Many of the studies in this survey present prototypes and proof-of-concepts that focus on collection and/or interpretation of data. For such prototypes to mature and be considered applicable on a population scale, there are likely application area-specific properties that need to be resolved within the following system engineering areas:

\subsection{Privacy and security}\label{ss:privacy}
Dietary data is generally not considered sensitive. 
However, if mined and analyzed, it might reveal highly private information on an individual's health and lifestyle situation.
For instance, a diet high in saturated fat and salt might indicate potential health problems that some subjects might want to keep private.
Dietary data should therefore be handled with a similar level of care 
for privacy and security as medical records. 
This is particularly important when it ends up in massive online data stores, along with other types of 
biometric data such as heart-rate measurements and activity logs.

Several applications already offer this kind of service for dietary data. 
For these applications, privacy and security are important and should be first-order design principles.
New privacy mechanisms that allow individuals to attach personalized privacy policies to raw data as well as aggregated and derived data may be necessary to secure ownership and control of dietary and health data that is collected by private parties~\cite{Johansen:2015:EPP:2797022.2797040}.

\subsection{Long term availability}
The time frame for availability of dietary data for analysis may depend on the application. To uncover whether certain health outcomes like, for example, cancer, are related to long-term dietary factors, the more historical data that is available for analysis the better. Since most health outcomes cannot be deterministically determined before they happen\footnote{With todays methods. This may change in the future.}, it is not known how long data should be available for analysis. 

Other applications, such as athletic performance diets or weight loss diets may have different requirements for data availability.

\subsection{Data federation and vendor lock-in}
Collecting dietary data for nutritional assessment can in many cases be a long-term project. Any data collection project spanning a long time frame may at some point face a situation where, for some reason, a need arises to switch data capture technology or data storage vendor. It may also become interesting to merge data from different sources. Proprietary data storage schemas and protocols make such changes problematic, making the process expensive. When research organizations perform studies on larger cohorts, regulations may demand that they do not grant providers of data gathering devices access to the gathered data. However, data storage protocols and schemas tend to be hard-coded into the devices, making it difficult or impossible to redirect storage to internal systems.  
Such situations are called \textit{vendor lock-in}.

Individuals doing self-intervention may have to deal with different devices with different properties and different analysis systems. For example, it may be useful to federate nutritional data with biometric data like heart rate, step count, weight etc. Frameworks such as Apple Health Kit,\footnote{https://developer.apple.com/healthkit/, Accessed May 2017} Google Fit,\footnote{https://www.google.com/fit/, Accessed May 2017} and Microsoft HealthVault\footnote{https://www.healthvault.com, Accessed May 2017} make such federation possible. However, there are still some applications and devices that do not support these federation services.

A common data schema for nutritional data will alleviate many of the purely technical issues regarding vendor lock-in and data federation. There may also be non-technical issues, where a vendor for example has a policy of not allowing data to be moved out of their systems. To avoid such situations, a general license framework for services gathering and storing data on behalf of individuals should be developed. This could be similar to the license schemes developed for open-source software, where different licenses offer different variations of data ownership and policies, and individuals may select which data license they want to use. It may also be advantageous to introduce a framework where trusted third-parties are used to certify that data vendors actually comply to license and policy agreements.

\subsection{Scalability}

We characterize the scalability issues of nutritional assessment using the \textit{Five V's of Big Data}:

\begin{description}
\item [Volume.] The surveyed methods collect varying amounts of data. The surveyors have not been able to investigate the exact amounts for each method, but there is reason to assume text-based methods and most sensor based methods collect limited amounts of data, compared to multimedia-based methods. Pictures, sound and video recordings represent significantly larger storage needs. Nutritional assessment sometimes incur following very large cohorts over long periods of time, which will collect large amounts of data regardless of the method. Nutritional assessment methods collecting large amounts of data need to consider where this data should be stored, when to synchronize client and server, and whether compression can be done losslessly. 
\item [Variety.] Few of the surveyed methods offer federation for holistic analysis and correlation with other sources. Those that do are typically mature applications that export data to external federation services, where data may be federated in the cloud. For scalability, local federation should be considered, but this may be hard to achieve unless data from each service is also stored locally.
\item [Velocity.] The methods that collect data non-stop at high rates tend to produce limited volumes as each data unit is relatively small. Most multimedia-based methods produce data when explicitly activated by the subject. However, some of the data is collected on the go, and thus data collecting devices may need buffering mechanisms to store data offline. Depending on the application, real-time analytics might be necessary. For example, users on strict diets may need instant feedback on whether or not to eat a specific meal in front of them, and how much of it. This type of analysis is not streaming, but very fast one-off analysis where some of the process might be offloaded to a remote server.  
\item [Validity.] We addressed the validity of collected data in our taxonomy in \t{cstable_tax}. 
\item [Volatility.] Depending on the type of intervention or investigation, nutritional assessments may span  different periods of time. \ac{24HR} requires detailed data collection for a fixed period of time, but may have to be stored and available for a much longer period in order to compare snapshots over time. Likewise, when large cohorts are investigated for large-scale nutritional trends it is important that data storage is non-volatile.   
\end{description}

\subsection{Personalization}
We have seen in \s{s:analysis} that a large variation in food choices causes challenges for machine learning models designed to recognize food items from images. It is likely, however, that each person does not choose at random from all possible foods every time that person eats. Adapting a model to persons or locations (e.g. restaurants) can thus help the recognition of food items by reducing the problem to choosing among a smaller range of foods, and thus also the scalability of such methods. It also makes automatic recognition of food more feasible for foods outside the pre-trained recognition model.

For those nutritional assessment methods that are designed to be used in interventions, personalization is key. Systems must adapt to the needs of each individual, and help individuals choose what is right for them. This might not only be a reactive method, data may also be used to proactively help make the right choices. 


\section{Conclusion}
This survey has reviewed the dominating techniques in nutritional assessment from an epidemiological standpoint. It has reviewed emerging fields within computer science and electrical engineering that all capture different aspects of eating and drinking with the aim of automating nutritional assessment. 

The motivation to automate nutritional assessment is mainly to: 
\begin{enumerate}
  \item remove self-reporting bias, and 
  \item reduce the effort to report, and 
  \item improve report quality, and 
  \item increase report quantity and detail.   
\end{enumerate}

As we have seen from \s{s:autom}, there are currently no known automated methods that remove self-reporting bias and reduce effort without sacrificing report quality, quantity and/or detail. The only methods that are completely without effort to report and still deliver nutritional data are those that bypass eating and instead detect biomarkers in blood. It is outside the scope of this survey to consider whether biomarker detection can replace highly detailed dietary data. 
However, it is likely that dietary data will always be necessary in order to be able to affect said biomarkers during an intervention.  

It is an open question whether it is possible to remove self-reporting bias completely. A free-roaming subject can choose to not take pictures of food items, and wearable sensors can be removed or turned off during, for instance, intake of alcohol. 

A relevant question is then whether incomplete data \textit{without} self-reporting bias is more worth than complete data \textit{with} self-reporting bias. It is possible that a lower level of detail can be useful, given that the data has been collected without bias. This is a particularly relevant point, as many of the methods that require little effort use pattern recognition in analysis of food intake. Work has been applied to recognize and classify food items from images, from combinations of images and recipes, from sensors capturing chewing and swallowing, and from sensors capturing upper body eating gestures. 

However, most prepared foods have ingredients like fats, minerals and sugars that are invisible to regular cameras. Other foods are completely homogenized, like soups and smoothies. Deriving detailed nutritional content from chewing and swallowing recordings is ongoing work. Methods combining user-images with online recipe databases are also ongoing work, but show promise to provide more details than images alone. In those methods where the quantity of each ingredient is known, as with digital nutrition scales, nutritional content of a meal can be calculated from standardized food databases, where the uncertainty of nutritional variations from, lets say, one apple to another, can be handled statistically. Thus, when exact nutritional content cannot easily be captured without analyzing the food itself, nutritional assessment may have to find techniques to deal with a lower level of detail.

While the different methods have different approaches to nutritional assessment, many aspects are similar. Automatic nutritional assessment has received significant adoption with the recent developments within the \textit{Internet of Things}, where wearable devices can be used to gather various biometric data. Most of the systems that apply wearable devices to record dietary intake without explicit user input have to make some deductions from raw data into data that can be used for nutritional assessment. This deduction process usually includes a number of data preparation steps, and some pattern recognition steps. Some have used \acp{SVM}, other have used neural nets with success. Thus, applied machine learning is one of the most important computer science topics within nutritional assessment.  

\section{Future work}
This survey has identified several concrete areas where further work is necessary, and that present interesting challenges for computer scientists: 
\begin{itemize}
  \item Create a standard data schema for structured data on dietary intake. 
  \item Create a standard data analysis framework for nutritional assessment.
  \item Determine what level of detail is necessary in different contexts, i.e. for weight loss, for optimal athletic performance, for chronic disease management, and so on.
  \item Determine whether recording of dietary intake can be bypassed, and the necessary biomarkers instead obtained directly from blood or other bodily fluids.
  \item Improve computer vision techniques that are used to recognize food. Here we refer to \cite{Meyers_2015_ICCV}, who made a listing of remaining challenges for computer vision techniques that are particularly exposed when trying to recognize food.
  \item Improve combined methods such as the neural joint embedding model that builds on text analysis of recipes and computer vision techniques to represent images, as well as other methods that combine different data sources.
  \item Properly evaluate any claimed improved nutritional assessment method to determine whether it actually benefits or advances the field or not. This must be done in collaboration with nutritional scientists in, for instance, randomized control studies.
\end{itemize}

%



%

 \section*{Acknowledgement}
This work was supported in part by the Norwegian Research Council, under grants 250138 (“Trans-Atlantic Corpore Sano”) and 263248/O70.

\bibliographystyle{ACM-Reference-Format}

\bibliography{diet_survey}


\end{document}